\documentclass[12pt]{article}

\usepackage{jneurosci}
\usepackage{graphicx}

\usepackage{times}

\newenvironment{sciabstract}{%
\begin{quote} \bf}
{\end{quote}}

\newcounter{lastnote}

\begin{document}
\begin{center}
{\LARGE The interplay between long- and short-range temporal correlations shapes cortex dynamics across vigilance states}\\
\vspace{0.8cm}
{\large Christian Meisel$^{1\ast}$, Andreas Klaus$^{1}$, Vladyslav V. Vyazovskiy$^{2}$ and Dietmar Plenz$^{1}$}\\
\end{center}
{\normalsize$^{1}$ Section on Critical Brain Dynamics, National Institute of Mental Health, Bethesda, Maryland 20892, USA}\\
{\normalsize$^{2}$ Department of Physiology, Anatomy and Genetics, University of Oxford, Parks Road, Oxford OX1 3PT, UK}\\
{\normalsize$^{\ast}$ corresponding author email address: christian@meisel.de}


\date{}

Short title: Cortical timescales in different vigilance states\\
Number of pages: 31\\
Number of figures: 4\\
Number of words for Abstract: 248\\
Number of words Introduction: 390\\
Number of words Discussion: 600\\
Conflict of Interest: N/A\\
Acknowledgements: This study was supported by the Intramural Research Program of the NIMH.  This study utilized the high-performance computational capabilities of the Biowulf Linux cluster at the National Institutes of Health, Bethesda, Md.\\



\baselineskip24pt



\newpage
\begin{sciabstract}
Increasing evidence suggests that cortical dynamics during wake exhibits long-range temporal correlations suitable to integrate inputs over extended periods of time to increase the signal-to-noise ratio in decision-making and working memory tasks. Accordingly, sleep has been suggested as a state characterized by a breakdown of long-range correlations; detailed measurements of neuronal timescales that support this view, however, have so far been lacking. 
Here we show that the long timescales measured at the individual neuron level in freely-behaving rats during the awake state are abrogated during non-REM (NREM) sleep. We provide evidence for the existence of two distinct states in terms of timescale dynamics in cortex: one which is characterized by long timescales which dominate during wake and REM sleep, and a second one characterized by the absence of long-range temporal correlations which characterizes NREM sleep. 
We observe that both timescale regimes can co-exist and, in combination, lead to an apparent gradual decline of long timescales during extended wake which is restored after sleep. 
Our results provide a missing link between the observed long timescales in individual neuron fluctuations during wake and the reported absence of long-term correlations during deep sleep in EEG and fMRI studies. They furthermore suggest a network-level function of sleep, to reorganize cortical networks towards states governed by slow cortex dynamics to ensure optimal function for the time awake.
\end{sciabstract}

\newpage
\section*{Significance Statement}
The lack of sleep deteriorates several key cognitive functions, yet the neuronal underpinnings of these deficits have remained elusive.
Cognitive capabilities are generally believed to benefit from a neural circuit's ability to reliably integrate information. 
Persistent network activity characterized by long timescales may provide the basis for this integration in cortex.
Here we show that long-range temporal correlations indicated by slowly decaying autocorrelation functions in neuronal activity are dependent on vigilance states. While wake and REM sleep exhibit long timescales, these long-range correlations break down during NREM sleep. Our findings thus suggest two distinct states in terms of timescale dynamics. During extended wake, the intermittency of short timescales can lead to an overall decline in cortical timescales.

\newpage
\section*{Introduction}

The ability to integrate information over time is a pre-requisite for brain function during wake.
Accumulating evidence suggests that the brain integrates information in a distinct progression of increasingly longer temporal timescales that closely follows the hierarchy of cortical organization \cite{Chaudhuri2015}.  
Evidence for such a hierarchical ordering and the existence of long timescales in higher order cortical areas has come from various experimental modalities including studies on non-human primates \cite{Ogawa2010,Murray2014}, human electrocorticography \cite{Honey2012} as well as fMRI \cite{Stephens2013}.
Functionally, these differentially expressed timescales provide an intriguing perspective on temporal information integration in the brain: by setting the duration over which a neural circuit integrates its inputs, longer times allow to increase the signal-to-noise ratio in decision-making and working memory \cite{Kiebel2008,Friston2012,Chaudhuri2015,Kringelbach2015}.

While these long timescales associated with long-range temporal correlations in the history of neural activity may provide a plausible mechanism for temporal integration during wake, their dependence on vigilance levels is currently less well understood. It has been suggested that during sleep the brain loses its ability to effectively integrate information across different cortical areas and across time, a functional loss that has also been proposed a potential mechanism for the concomitant loss of consciousness \cite{Tononi2008}. Indeed, a decrease in information integration during nonrapid eye movement (NREM) sleep has been observed in space, quantified by the effective connectivity between different cortical regions \cite{Massimini2005}, and time \cite{Tagliazucchi2013}. However, these studies have focused on large-scale brain signals measured by EEG and fMRI; their link to the long timescales governing individual neuron activity is missing.   
Apart from the effect of sleep on cortical timescales, it is conceivable that also more subtle vigilance changes may impact the timescales of cortical dynamics. For example, cognitive capabilities and the ability to process information are known to be affected by extended waking \cite{VanDongen2003,Banks2007,Mignot2008,Killgore2010}.
The impact of extended wakefulness on cortical timescales, however, is currently unknown. 

Here, we sought to systematically characterize the timescales governing cortical dynamics during wake, extended waking, and its dependency on sleep. We hypothesized that the breakdown in long-range temporal correlations during sleep will be visible at the level of individual neuron spiking activity.
For this purpose, we analyzed neuron activity at the individual and multi-unit level along with local field potentials (LFPs) across different cortical regions in freely-behaving rats.

\section*{Materials and Methods}
We investigated two datasets where cortical activity was monitored across different vigilance states in rats. In the first dataset, local field potentials (LFP) and multi-unit activity (MUA) were recorded from frontal cortex during 6 hours of sleep deprivation and after a consecutive recovery period (n=7 animals; dataset 1). In the second dataset, LFP and single-unit activity were recorded from frontal and parietal cortex areas during 4 hours of sleep deprivation and during consecutive recovery sleep (n=13 animals; dataset 2).

\subsection*{Dataset 1}
Adult male rats (Spraque Dawley, total n=7) were used in this dataset. All rats were housed individually in transparent Plexiglas cages. Lighting and temperature ($21\pm0.5 C$) were kept constant (LD 12:12, light on at 6 am, food and water available ad libitum).

\subsubsection*{Implantation of microelectrode arrays}
Procedures were in accordance with National Institutes of Health guidelines. Animal procedures were approved by the National Institute of Mental Health Animal Care and Use Committee.
At age 4-6 weeks, multichannel microelectrode arrays were implanted in the right frontal cortex (B: +1-2 mm, L: 2.5-3.5 mm) under deep isoflurane anesthesia ($1.5-2 \%$ volume) and presence of the analgesic ketoprofen (5 mg/kg, subcutaneous). Ketoprofen was given for up to two days post-surgery and animals were allowed to recover for 5 days before recordings.
The arrays had 32 channels (8x4, $200 \mu m$ inter-electrode spacing; $23 \mu m$ electrode diameter; Neuronexus, Ann Arbor, MI, USA).
The surgical procedure was performed in sterile conditions where, first, a $\sim$ 2x2 mm craniotomy was made using a high-speed surgical drill. After an incision in the dura the electrode array was then lowered into the brain tissue by penetrating the pia mater. Electrodes were slowly advanced further into brain tissue until all rows of the array were at the level of superficial cortical layers (layers 2/3) and most channels showed robust single- or multiunit activity. After application of silicon gel to seal the craniotomy and to protect the surface of the brain, dental acrylic was placed around the electrode array to fix it to the skull. The ground wire was connected to a scull screw located 1 mm posterior to bregma.

\subsubsection*{Experimental design}
At least 5 days were allowed for recovery after surgery. Sleep deprivation began at light onset (6 am) and involved continuous observation of the animal and its LFP recording. Animals were sleep deprived for a total of 6 hours after which an undisturbed recovery period of 5-6 hours duration followed. 
Prolonged wakefulness was achieved by providing rats with novel objects and/or delivering low-level auditory or visual stimuli. Prior to the experiment, rats were habituated to the experimenter and to the exposure to novel objects (including nesting and bedding material from other rat cages, tubes of different shape and color, little colored blocks, rubber balls and boxes of different shapes and colors). Care was taken not to stress animals; they never appeared to engage in freezing or aggressive behavior, and exhibited no signs of discomfort.

\subsubsection*{LFP and MUA recording during prolonged wakefulness}
Neuronal activity including LFP and extracellular unit activity was recorded at 30 kHz and stored for offline analysis (Cerebus, Blackrock Microsystems). At least 20 minutes of continuous LFP and MUA were recorded for each hour and experiment during prolonged wakefulness (from 0 to 6 hours, corresponding to SD0 to SD6, respectively) as well as after the recovery period (REC). It has previously been pointed out that wake is not a homogenous state, and that it is therefore important to perform analyses within substates (quiet and active wake) as similar as possible \cite{Fisher2016}. We here, similar to previous work \cite{Vyazovskiy2011}, focused on "quiet wake" (Q) episodes, i.e. periods when the animal is alert and has its eyes open, readily responds to stimuli while maintaining vigilance and posture, but is immobile. To find these episodes, we first marked periods when the animal behaviorally appeared to be in a Q state, i.e. had its eyes open, readily responded to stimuli, maintained vigilance and posture, but was immobile. This included the removal of episodes with rearing, grooming, eating or drinking. Channels containing artifacts or noise were removed from further analysis of LFP and MUA. We then carefully selected time periods of at least 4 s duration that showed LFP signals characterized by typical wake activity, i.e. LFP signals characterized by mainly low voltage activity with theta waves and no extended periods of slow waves typical for sleep ($\geq\sim$ 1 s). 

Similar to previous reports \cite{Vyazovskiy2011} we observed occasional large-amplitude activity in these select periods of Q, which typically presented as single positive half-waves in the frequency range from 2-6 Hz, but could occasionally also appear in succession of 2 or more waves (Fig. \ref{fig_1}). Events like these are often not detectable in EEG \cite{Vyazovskiy2011} and are therefore part of what is traditionally considered the wake state. To test if or to what extent our results depended on the occurrence of these events, all analyses were repeated on data where segments with this kind of slow wave activity had been eliminated (see below). As a further control, such large-amplitude segments were carefully removed from analyses in dataset 2 (see below). 

Multi-unit spikes were extracted from the high-pass filtered signal ($\geq 250 Hz$) by applying a threshold at negative 5 times the root mean square of the signal using the Cerebus Central software (Blackrock Microsystems). Thresholded waveforms were subsequently offline-sorted using the Offline Spike Sorter (Plexon Inc., Dallas, TX, USA) to eliminate artifactual waveforms caused by electrical or mechanical noise.
Principal components (PCs, set to 3) were extracted and automatic cluster selection was performed using the T-Distribution Expectation-Maximization (E-M) method \cite{Shoham2003}. Only electrodes and units were used for analysis for which the typical biphasic waveform was evident. LFPs were first downsampled to 500 Hz and band-pass filtered (1-40 Hz) using a phase neutral filter by applying a second order Butterworth filter in both directions. A notch filter to eliminate line noise was applied subsequently. 

\subsection*{Dataset 2}
For dataset 2, we used recordings from a previous study described in detail in \cite{Vyazovskiy2011}. In short, data consisted of male WKY rats from which LFP and unit activity were recorded in frontal (n=11) and/or parietal (n=9) cortex with 16-channel micro-arrays. Similar to dataset 1, rats were housed individually in transparent Plexiglas cages (light:dark 12:12 h, light on at 10:00; food and water ad libitum and replaced daily at 10:00). 
Data recording and online spike sorting were done with the Multichannel Neurophysiology Recording and Stimulation System (Tucker-Davis Technologies Inc). Unit activity was collected continuously (25 kHz, 300--5,000 Hz), concomitantly with the LFPs from the same electrodes and epidural EEGs (both 256Hz, 0.1--100 Hz). Data consisted of 4 hours of sleep deprivation and 6 hours of consecutive recovery. In each animal, two to four experiments with 4 h of sleep deprivation were performed (at least 5 days apart). Sleep stages were scored offline by visual inspection of 4-s epochs. Spike sorting was performed by principal component analysis (PCA) followed by split and merge expectation maximization (SMEM) clustering algorithm. 
Individual unit activity was binned in 2-ms bins and stored for further offline analysis. All analyses in this study were performed on artifact-free 4s-intervals classified as vigilance stages pertaining to either quiet wake (Q), REM sleep (R) and NREM sleep (N). Quiet wake epochs were defined as in dataset 1 (see above) while, additionally, carefully excluding also short epochs with large-amplitude LFP activity. For further details regarding these data, see \cite{Vyazovskiy2011}.

\subsection*{Spike-count autocorrelation}
The temporal autocorrelation of spike counts was computed for MUA (dataset 1) and putative single units (dataset 2) similar to previous work \cite{Ogawa2010,Murray2014}. The 4s-intervals were divided into separate, successive time bins of duration $\Delta$ which we set to $\Delta=100ms$. Across different 4s-intervals within the same animal and hour during sleep deprivation (datasets 1 and 2), during recovery sleep (dataset 2) and after recovery sleep (dataset 1), we calculated the correlation between spike counts N in two time bins, indexed by their onset times $i\Delta$ and $j\Delta$, using the Pearson’s correlation coefficient R:

\begin{equation}
\label{pearson}
R(i\Delta,j\Delta)=\frac{\mathrm{Cov}(N(i\Delta),N(j\Delta))}{\sqrt{\mathrm{Var}(N(i\Delta)) \times \mathrm{Var}(N(j\Delta))}}
\end{equation}

where covariance (Cov) and variance (Var) are computed across 4s-intervals for those time bins. To allow reliable computation of spike-count autocorrelation, we required that each (multi-)unit fired at least 8 times during the 4s-interval and that at least 10 4s-intervals within each animal and hour were available \cite{Ogawa2010}; similar to previous work this threshold level was not essential to observe our main findings. One rat of dataset 1 did not exhibit MUA at all time points during the sleep deprivation period and was thus excluded from the analysis, so that spike-autocorrelation was derived from a total of n=6 rats in dataset 1. The decay of autocorrelation was fit by an exponential decay to the population of neurons, i.e. the mean across autocorrelation functions from all units, for each time point and rat:

\begin{equation}
\label{expdecay}
R(k\Delta)=A(e^{-\lambda k\Delta} + B),
\end{equation}

where $\lambda$ is the decay rate. Fitting was done using the \textsc{Python} (Python Software Foundation, version 2.7) function \textsc{scipy.optimize.curve fit}. As a minimal goodness of fit we required $R^2\ge0.5$ which excluded some rare outliers due to poor fits.

\subsection*{Recovery from large-activity events}
MUA activity increased with negative LFP (nLFP) excursions (Fig. \ref{fig_2}). To analyze how MUA and LFP activity recovered from such large-activity events in dataset 1, we identified these events of increased MUA activity in each channel and for each 4s-interval as nLFP excursions greater or equal -2 standard deviations (SD) after z-transformation (Fig. \ref{fig_2} a). For each such nLFP event, both the LFP as well as the MUA in this channel for the time $\pm 100 ms$ around the most negative LFP excursion was then saved for further analysis. We required the next consecutive nLFP event to be at least 200 ms away. Next, the recovery from these nLFP was analyzed. nLFP segments were averaged for each time point (hour during sleep deprivation or after recovery period) and animal (Fig. \ref{fig_2} b) and fit by an exponential decay to derive recovery rates for each time point and each animal. The recovery rates $\lambda$ were derived by fitting an exponential function to the nLFP:

\begin{equation}
\label{expdecay2}
S(t)=Ae^{-\lambda t} + B
\end{equation}

using the \textsc{Python} (Python Software Foundation, version 2.7) function \textsc{scipy.optimize.curve fit}. The recovery was fit from 30 ms after the most-negative LFP deflection to the local maximum within 100 ms after the most-negative LFP deflection since LFP sometimes exhibited a bleed-through of the spike trace in the vicinity of the most negative excursion \cite{Ray2008,Ray2011,Nauhaus2009,Nauhaus2012}. To avoid distortion by this distinct wave, we chose to fit the LFP recovery from large-activity events at 30 ms after the most-negative deflection.

\subsection*{Detection of bimodality in 2-6 Hz half-waves}
For this analysis, LFP signals were further band-pass filtered (2-6 Hz) using a phase neutral filter by applying a second order Butterworth filter in both directions in each 4s-interval. The histogram of positive 2-6 Hz half-waves exhibited a clear bimodality in double-logarithmic plot when the occasional brief large-amplitude activity events were not removed (dataset1; see above; Fig. \ref{fig_1} c). 

In dataset 1, to test the dependence of our results on the occurrence of these events, all analyses were repeated for segments with or without this kind of large-amplitude slow wave activity. Specifically, we determined whether the maximum of the 2-6 Hz filtered LFP signal in each channel and 4-s segment was larger than a given threshold, and then separately analyzed spike autocorrelation and nLFP recovery for units and LFP data based on whether there was an event larger than this threshold or not. We set the threshold individually for each channel and each 4-s segment as the median plus one standard deviation of the positive 2-6 Hz half-waves (Fig. \ref{fig_1} c, vertical magenta line). This threshold reliably removed all large-amplitude events (i.e. the bimodality in the LFP amplitude distribution) while maintaining a large-enough number of data segments for analysis both above and below threshold.

In dataset 2, segments during quiet-wake (Q) were carefully chosen not to contain any of these high-amplitude LFP events. This is evidenced by the 2-6 Hz LFP amplitude distribution which did not exhibit a prominent bimodality (Fig. \ref{fig_3} d).

\subsection*{Computational model}
To develop better insights into the determinants of timescale changes in cortical network acitivty, we studied a neuron network model that could exhibit long timescales associated with long-range temporal correlations. The neuron network model consisted of N=100 all-to-all coupled, binary-state neurons with the following dynamical rules: If neuron $j$ spiked at time $t$ (i.e., $s_j(t)=1$) as part of an active set of neurons $J(t)$, a postsynaptic neuron $i$ will spike at time $t+1$ according to

\begin{equation}
\label{model}
s_i(t+1)=\Theta [1 - \prod_{j\in J(t)}(1-p_{ij}) - \zeta(t)]
\end{equation}

where $\Theta[x]$ is the unit step function and $\zeta(t)$ is a random number from a uniform distribution on [0,1] to account for the probabilistic nature and variability of unitary synaptic efficacy.
$p_{ij}$ are the asymmetric synaptic coupling strengths between each pair of neurons which are first drawn from a uniform distribution on [0,1] and then multiplied by $N \cdot K /\sum p_{ij}$. The parameter $K$ is consequently related to the average connectivity and is therefore a control parameter for the dynamics of the network: at $K=1$ each spiking neuron excites, on average, exactly one postsynaptic neuron meaning the network is critical. 
Conversely, at $K<1$ activity dies out prematurely and the system is subcritical; at $K>1$ each neuron excites on average more than on postsynaptic neuron and the system is supercritical \cite{Shew2009}. 
We instantiated the model at different average connectivies $K=1$ to study the spike autocorrelation and recovery from large-activity events in similar ways as in the experimental data. Network activity was introduced by randomly setting one neuron to active at each time step.

The temporal autocorrelation of neurons was studied by simulating network activity for 5050 timesteps. After the initial 50 timesteps during which network activity typically reached a stable level, neuron activity was binned in 50-timestep bins, resulting in spike-count timeseries of length 100 for each neuron. This process was repeated for 100 network simulation runs which allowed the calculation of Pearson’s correlation coefficients R (eq. \ref{pearson}) for different time bins and across simulation runs analogously to the experimental data. The population autocorrelation function was, also analogously to experimental data, derived by averaging across all neurons and consequently fit by an exponential decay (eq. \ref{expdecay}) to derive a recovery rate $\lambda$.

To study the recovery from large-activity network events relevant to our experimental analysis, the recovery from large, intrinsic fluctuations in network activity was monitored. We simulated network activity for 5050 timesteps. Network activity typically reached a stable level after about 50 timesteps. From the following 5000 timesteps, we identified large intrinsic fluctuations in the spike histogram greater than a threshold $T$ which we set to $T=2$ standard deviations. Once a large fluctuation of network activity, i.e. in the spike histogram, had been identified, we required the following fluctuation event to be at least 50 timesteps apart. We averaged over all fluctuation events from a total of 100 network simulation. We fit an exponential decay (eq. \ref{expdecay2}) to these averages to derive a recovery rate $\lambda$.

To implement offline periods of spiking similar to those observed during sleep deprivation, the activity of an individual neurons was set to zero with a probability $p_{\rm{OFFLINE}}$ at each time step during the simulation.

\section*{Results}

We first assessed the timescales governing cortical dynamics at the individual neuron level in three different vigilance states: REM sleep, NREM sleep and quiet wake. 
For each vigilance state, the spike-count autocorrelation was calculated as the correlation coefficient between the number of putative single-unit spikes in pairs of time bins across all artifact-free 4s-segments. Following previous work \cite{Ogawa2010,Murray2014}, we derived an autocorrelation function reflective of the population- or network-level activity in each animal by averaging functions across units (Fig. \ref{fig_1} a). We observed slow autocorrelation decays indicative of long-range temporal correlations during quiet wake in frontal cortex. 
While REM sleep exhibited a similar, slowly decaying autocorrelation function, NREM sleep differed profoundly, exhibiting a fast decaying autocorrelation function at significantly higher rate (ANOVA and post hoc unpaired t-test; Fig. \ref{fig_1} a, b).
To assure that these timescale differences were independent of firing rates, we repeated the analyses using only neurons with average firing rates between 3 and 5 Hz in all vigilance states. While this selection abolished all firing rate differences between vigilance states, the autocorrelation decay differences were maintained (Fig. \ref{fig_1} c; ANOVA, F=29, p=4e-13). 

Next, we extended our analysis to recordings from parietal cortex to investigate how universal the vigilance-dependent timescale differences were across cortex. No statistically significant difference between frontal and parietal areas was observed when all quiet wake segments were compared (frontal: $0.0017\pm0.0002$; parietal: $0.0015 \pm0.0002$; p=0.39; unpaired t-test). In absolute terms, these values correspond to timescales of about 400 ms, which is in the order of magnitude also observed in areas of higher cortical hierarchy in non-human primates \cite{Murray2014}. Similar to frontal cortex, we observed a much more rapid spike-count autocorrelation decay during NREM sleep in comparison to quiet wake and REM sleep indicative of a loss of long-range temporal correlations (Fig. \ref{fig_1} d--f). 

These results indicate distinct timescales in different vigilance states: long timescales associated with a slow autocorrelation decay during quiet wake and REM sleep, and short timescales during NREM sleep. 
We next investigated the persistence of timescales within vigilance states and across time. 
Previous work has shown that wake is not a homogeneous state. Spontaneous waking is associated with more pronounced slow rhythms in the EEG \cite{Finelli2000,Strijkstra03}, changes in firing rates \cite{Fisher2016} and the occurrence of local OFF periods where neurons stop firing and go "offline" \cite{Poulet2008,Vyazovskiy2011,McGinley2015}. Some of these changes, like local offline periods, may not show in scalp EEG \cite{Vyazovskiy2011}; the underlying state may thus still be considered wake in the traditional sense. We therefore studied neuron timescale dynamics in wake by systematically taking the effect of offline periods into consideration. 

We first analyzed multi-unit activity of frontal cortex in freely-behaving rats during 6 hours of sleep deprivation and after a consecutive recovery period (dataset 1). 
In these data, the spike autocorrelation analysis revealed a progressive increase of decay rates during sleep deprivation reflective of the faster autocorrelation decay (ANOVA, F=2.91, p=0.015; Fig. \ref{fig_2} a, b). Consecutive sleep, conversely, recovered the slower rates. Post hoc statistical comparison of beginning (SD0) to end of sleep deprivation (SD6), and end of sleep deprivation (SD6) to post recovery sleep (REC) revealed both changes to be significant (p=0.002 and p=0.01, respectively; two-tailed paired t-test). 
Similar to previous work \cite{Vyazovskiy2011}, we observed occasional large positive LFP half-waves in the 2-6 Hz frequency band that were accompanied by pauses in the MUA (Fig. \ref{fig_2} c). When plotted in double logarithmic coordinates, the amplitude of positive half-waves in this frequency band exhibited a distinct bimodality in its density distribution (Fig. \ref{fig_2} c). Such a bimodality has also been observed in other frequency bands, where it was taken as indication for the existence of two co-existing different states between which dynamics can switch \cite{Freyer2011}. In our case, the observed bimodality may similarly suggest two distinct states characterized by different low-frequency power levels. We hypothesized that large-amplitude LFP events might correspond to a more sleep-like state and thus potentially imprint shorter timescales. We thus repeated our analysis in select data with or without any such events. Specifically, we defined a threshold for each animal and channel individually, that separated the two peaks in the LFP amplitude distribution (Fig. \ref{fig_2} c right, magenta line). We then separately analyzed spike-count autocorrelations in all channels where amplitudes in the 2-6 Hz frequency band either exceeded this threshold at least once or stayed below it throughout a 4-s segment. 
The separation of sleep deprivation data along these two peaks in the LFP distribution revealed that timescale changes were only apparent in segments that contained high-amplitude LFP events. Timescales did not change during sleep deprivation in the low-amplitude regime (ANOVA; Fig. \ref{fig_2} d, e).
These observations indicate distinct dynamics with respect to the two peaks in the LFP distribution. The intermittency of both dynamical states leads to the apparent shortening of intrinsic timescales characterizing spontaneous fluctuations of neuron activity during sustained wakefulness.

To confirm the causal role of intermittent large-amplitude LFP events for the apparent timescale decline during extended wake, we repeated our analysis in dataset 2 where such large-amplitude LFP events had been excluded altogether in the selection of quiet wake data (as confirmed by the absence of a distinct bimodality in the LFP amplitude distribution; Fig. \ref{fig_2} f, blue). Consequently, autocorrelation decays obtained from quiet wake exhibited no statistically significant change during sleep deprivation in this dataset (frontal and parietal; SD0 compared to SD4; unpaired t-test; Fig. \ref{fig_2} g). During extended wake, the occurrence of shorter timescales is thus linked to the presence of intermittent large-amplitude LFP events.
Over the course of NREM sleep, autocorrelation decays became significantly less steep in frontal cortex (ANOVA, F=3.84, p=0.003; Fig. \ref{fig_2} i). A similar trend was visible in parietal cortex although it did not reach significance level there. REM sleep, conversely, did not exhibit a significant change in any region (Fig. \ref{fig_2} h). The differential changes observed over the course of NREM sleep but not REM sleep thus provide further evidence for the relevance of low frequency large-amplitude events as the essential mechanism for the apparent timescale change. Low frequency large-amplitude events were present during NREM but not REM sleep (Fig. \ref{fig_2} f, brown vs. green line), and are known to decrease over the course of NREM sleep \cite{Borbely1981,Finelli2000,Vyazovskiy2011} which can account for the less steep spike autocorrelation decays observed.

There exists an intimate relationship between the autocorrelation function governing spontaneous fluctuations and the time it takes a signal to recover from small perturbations \cite{Wissel1984,Ives1995,Scheffer2009}.
From this dynamical systems perspective, a slow autocorrelation decay also implies a slow recovery from perturbations or intrinsic fluctuations. We hypothesized that equivalent wake-time dependent changes quantified by autocorrelation decays should also manifest in recovery rates from intrinsic fluctuations.
We first identified sizeable negative excursions from the mean in each LFP channel (Fig. \ref{fig_3} a). Next, we averaged across all these LFP segments for each rat and at each time point (dataset 1) to quantify the return to baseline activity during sleep deprivation and after consecutive recovery sleep. Negative LFP averages typically coincided with increased spiking activity (Fig. \ref{fig_3} b, c). 
To quantify the recovery rate by which LFPs returned to baseline after a negative excursion, we fit the return with an exponential function. Recovery rates obtained from these fits increased during sleep deprivation and reversed to lower values after consecutive sleep (ANOVA, F=3.5, p=0.004; Fig. \ref{fig_3} d, e). Post hoc statistical comparison between the beginning (SD0) and the end of sleep deprivation (SD6), and between the end of sleep deprivation (SD6) and after recovery sleep (REC) revealed both changes to be significant (p=0.002 and p=0.01, respectively; two-tailed paired t-test). We hypothesized that large-amplitude LFP events would again drive the observed timescale decline. As with spike-count autocorrelation, we thus repeated our analysis for both the low-amplitude and the high-amplitude regimes separately. In agreement with our observations from spike-count autocorrelations, timescales changes were only apparent in segments also containing high-amplitude LFP events, while timescales did not change during sleep deprivation in the low-amplitude regime (ANOVA; Fig. \ref{fig_3} f, g).

Together, results indicated that intermittent periods of large-amplitude LFP can lead to an apparent decline of long cortical timescales during wake. Our observations also suggested that these LFP events are characterized by pauses in neuron spiking. 
To test whether these intermittent offline periods were sufficient to account for timescale changes, we studied a neuron network model exhibiting long timescales associated with long-range temporal correlations analogous to experiment. Specifically, the network's average connectivity was set so that network dynamics was located right at the transition between an inactive and an active network phase \cite{Shew2009}. At this transition state, the network exhibited long timescales associated with both the autocorrelation decay and the recovery from intrinsic fluctuations. We then implemented the possibility that neurons go "offline" for a short amount of time with a certain probability. We observed that this ability to go offline was sufficient to reproduce our experimental findings during wake.
In particular, with higher probability for neurons to go offline, faster declines in spike-count autocorrelation and recovery from intrinsic fluctuations were observed (Fig. \ref{fig_4}). These modeling results thus suggest that an increasing tendency for neurons to go offline by themselves (e.g. during sleep deprivation) or in synchrony (e.g. during NREM sleep) is sufficient to account for the shortened timescales.

\section*{Discussion}

The ability to integrate information over time is a pre-requisite for brain function during wake. Information integration is believed to be supported by long timescales associated with long-range temporal correlations in neuron spiking activity. We here provide first-time evidence at the neuronal level that these timescales are abrogated during NREM sleep. In contrast, quiet wake and REM sleep are characterized by similar, long timescales. 
Our results provide indication for the existence of distinct states in terms of timescale dynamics: one, which is characterized by long timescales that dominate during wake and REM sleep, and a second one characterized by the absence of long timescales predominantly observed in NREM sleep. 
Our sleep deprivation results suggest that transitions between these two states can occur rapidly and intermittently which, on average, can lead to an apparent progressive decline of timescales.

The present work provides a missing link between long cortical timescales \cite{Ogawa2010,Honey2012,Stephens2013,Murray2014,Chaudhuri2015} to observations indicating disrupted long-range correlations during NREM sleep \cite{Massimini2005,Tagliazucchi2013}. The observed breakdown of long timescales in ongoing spiking activity here provides insights that complements these reports based on large-scale signals measured by EEG and fMRI.
From a theory point of view, the timescales quantified by the autocorrelation of intrinsic fluctuations are intimately linked to the timescales that control the response to external perturbations \cite{Wissel1984,Ives1995,Scheffer2009,Meisel2015}. The fast autocorrelation decay observed in our data during NREM sleep thus links to observations of reduced stimulation responses using TMS in the same vigilance state \cite{Massimini2005}.

In detailed analyses of the apparent gradual timescale decline during extended wake, we identified intermittent transitions between two different states. We found that the state characterized by a fast decaying autocorrelation increasingly dominated the overall dynamics, while autocorrelation decay rates did not change for the low-amplitude state during sleep deprivation. The two peaks visible in the LFP amplitude distribution thus point to distinct dynamical states, governed by long timescales in the low amplitude regime, and shorter timescales in the high amplitude regime. The identification of two different states being at play relied on the ability to record data at high temporal and spatial resolution in our data. It is therefore possible that the gradual decline in timescales observed during the transition into NREM sleep \cite{Tagliazucchi2013} is similarly the result of two different co-existing states which a more coarse method such as fMRI may not be able to resolve in the temporal domain. Several subcortical structures in the brainstem, hypothalamus and basal forebrain are known to regulate the maintenance of waking and sleep states through neuro-modulatory action \cite{Jones2005}. It is conceivable that these subcortical structures control the transient occurrences of the large-amplitude LFP epochs observed in some of our data during wake.

Functionally, the shorter timescales during NREM sleep indicate a severely reduced ability to integrate information. During sleep deprivation, shorter timescales induced by intermittent, sleep-like, high-amplitude LFP events may be similarly detrimental for cortical networks to process information. As such, this decline may provide a perspective on several cognitive capabilities known to decrease during extended wake \cite{VanDongen2003,Banks2007,Mignot2008,Killgore2010,Lo2012}. In fact, a two-state dynamics, as observed here, may help to explain some at times controversial findings with respect to the specifics performance impairments. Although it is well-established that performance, for example quantified in the psychomotor vigilance test (PVT), decreases, on average, as a function of time awake, a subject may still perform almost normally even after extended periods of sleep deprivation during individual trials in the PVT \cite{Banks2007}. It is conceivable that normal performance could still be achieved at times when dynamics is not compromised by intermittent high-amplitude events, while performance will decrease on average due to the growing dominance of intermittent epochs with disrupted timescales during sleep deprivation.

Long-range temporal correlations are generic features that arise when a system is in the vicinity of a critical state \cite{Bak1987}.
Mounting evidence from computational and experimental studies indicates that cortical neural networks operate at some sort of critical state \cite{Bornholdt2000,Beggs2003,Haldeman2005,Shew2009,Meisel2012,Markram2015}. The observation of long timescales during wake is in agreement with this criticality hypothesis \cite{Linkenkaer2001}. Conversely, fast decaying autocorrelation functions such as during NREM sleep and also during sleep deprivation may be seen as indication for a dynamical state where critical dynamics is disrupted. In a previous study based on EEG, we reported fading signatures of critical dynamics during sleep deprivation \cite{Meisel2013}. Our results here demonstrate an increasing disruption of long timescales during sleep deprivation at the individual neuron level which complements these earlier findings. From this perspective, our observations support an intriguing hypothesis for a function of sleep, to re-organize cortical networks towards critical dynamics with long-range temporal correlations for optimal functioning during wake.




\clearpage


\section*{Acknowledgements}
This study was supported by the Intramural Research Program of the NIMH.  This study utilized the high-performance computational capabilities of the Biowulf Linux cluster at the National Institutes of Health, Bethesda, Md.

􏴎\newpage

\begin{figure}
\centering
\caption{\label{fig_1}
Breakdown of long-range temporal spiking correlations during NREM sleep. a, Spike-count autocorrelation (AC) decays rapidly in NREM sleep while other vigilance states have similar slow decays. Mean AC over animals (n=13; s.e.m.; single-unit activity; frontal cortex; quiet wake, Q, REM sleep, R, NREM sleep N). b, Summary of decay rates. Decay rates obtained from exponential fits over the whole autocorrelation function. c, Changes in autocorrelation decays are independent of differences in spiking rate between vigilance states. 
d--f, Breakdown of long-range temporal spiking correlation in NREM sleep in parietal cortex. Equivalent analysis as in a--c.}
\end{figure}

\begin{figure}
\centering
\caption{\label{fig_2}
Two-state dynamics underlies the decline in long-range temporal correlations during sleep deprivation. a, Decay in spike-count autocorrelation (AC) accelerates as a function of time awake and recovers after sleep (multi-unit activity; quiet wake; 6 hours of sleep deprivation SD0 to SD6 followed by 5--6 hr recovery sleep, REC). Mean AC over all multi-units (n=26; 1 rat; s.e.m.). Solid lines: exponential fit. Left: linear plot. Right: double-logarithmic plot. SD3 line slightly shifted upwards for visibility. b, Summary of decline in decay rates and recovery for all n=6 animals (s.e.m). c, Identification of two-state dynamics underlying decay in long-range temporal correlations with sleep deprivation. Large-amplitude LFP events are intermittently observed and typically coincide with pauses in multi-unit activity (top; offline periods marked as gray areas in zoomed inset). These LFP events typically present as one or more positive half-waves predominantly in the frequency range 2-6 Hz (bottom, wavelet spectrogram and positive LFP trace bandpass filtered between 2-6 Hz). Right: Bimodality in the histogram of positive 2-6 Hz waves. The bimodality was used to separate the large-amplitude LFP containing putative offline periods from other activity for each rat individually (magenta line). d, e, Decline in long-range temporal correlations is restricted to data segments containing large-amplitude LFP events.
f, Histogram of positive LFP 2-6 Hz waves sorted by vigilance states (dataset 2; top row frontal, bottom row parietal).
g, Autocorrelation decays remains unchanged (right) in data without large-amplitude LFP events during sleep deprivation (quiet wake; 4 hours of sleep deprivation SD0, SD4).
h, Autocorrelation decay rates during REM sleep remain unchanged (dataset 2; 6 hour recovery sleep; R1-R6).
i, Autocorrelation decay rates in NREM sleep decrease over the 6 hour recovery sleep (dataset 2; N1-N6).
}
\end{figure}

\begin{figure}
\centering
\caption{\label{fig_3}
Two-state dynamics underlies the accelerated recovery from large-activity events during sleep deprivation. 
a, Identification of large-activity events from LFP. Negative LFP (nLFP; <-2 standard deviations) deflections were averaged over each time point during the sleep deprivation experiment (SD0 to SD6, REC) and for each rat. b, Mean nLFP over all events and channels (1 rat; s.e.m.). c, Corresponding nLFP-triggered spike histogram. The histogram is normalized to mean activity from -150 to -100 ms.
d, Recovery from intrinsic nLFP excursions accelerates as a function of time awake and recovers after sleep.
Mean nLFP traces from the beginning and end of sleep deprivation from one rat. e, Summary of nLFP recovery rates for all n=7 animals (s.e.m.).
f, g, More rapid recovery from large-activity events is restricted to data segments containing large-amplitude LFP events.
}
\end{figure}

\begin{figure}
\centering
\caption{\label{fig_4}
Neuron network model with the ability of neurons to go offline captures the differential decline of long-range temporal correlations. The network was instantiated at a connectivity K=1 and with different probablities $p_{\rm{OFFLINE}}$ for neurons to go offline for a brief amount of time. a, Raster plot without the presence of offline periods. b, Raster plot with offline periods($p_{\rm{OFFLINE}}=0.1$, grey shaded areas). c, Autocorrelation functions for spiking activity in a network without offline periods ($p_{\rm{OFFLINE}}=0$, blue) and in a network with offline periods ($p_{\rm{OFFLINE}}=0.3$, red) in neuron activity. d, Autocorrelation decay rates increase with higher probability for offline periods. e, Recovery from intrinsic fluctuations ($p_{\rm{OFFLINE}}=0$, blue; $p_{\rm{OFFLINE}}=0.3$, red). f, Recovery rates from intrinsic fluctuations increase with higher probability for offline periods.
}
\end{figure}

\clearpage

\end{document}